\documentclass[a4paper]{article}
\usepackage[utf8]{inputenc} 

\usepackage{todonotes}
\setuptodonotes{inline}

\usepackage{url}  
\usepackage{graphicx}

\usepackage{algorithm}
\usepackage{algorithmicx}
\usepackage[noend]{algpseudocode}
\floatname{algorithm}{Algorithm}

\newcommand{\bspq}[0]{BspQI}
\newcommand{\bsss}[0]{BssSI}

\author{Przemysław Stawczyk and Robert Nowak\\
  Institute of Computer Science, Warsaw University of Technology,\\
  Warsaw, Poland}

\title{Fast genomic optical map assembly algorithm using binary representation}

\date{2022.10.12}

\begin{document}

\maketitle

\footnote{Further author information: (Send correspondence to Robert Nowak)
  Robert Nowak: Tel: +48 22 2347718; E-mail: robert.nowak@pw.edu.pl}

\begin{abstract}
  Reducing the cost of sequencing genomes provided by next-generation sequencing technologies has greatly increased the number of genomic projects.
  As a result, there is a growing need for better assembly and assembly validation methods.
  One promising idea is to use heterogeneous data in assembly projects.
  Optical Mapping (OM) is beneficial in validating genomic assemblies, correction and scaffolding.
  Single raw OM read describes a DNA molecule's long fragment, up to 1Mbp.
  Raw OM data from the same genome could be assembled to create consensus maps that span an entire chromosome.
  The assembly process is computationally hard because of the large number of errors in input data.
  This work describes a new algorithm and computer program to assemble OM reads without a reference genome.  In our algorithm, we explored binary representation for genome maps.  We focused on the efficiency of data structures and algorithms and scale on parallel platforms.  The algorithm consists of several steps, of which the most important are : (1) conversion of the restriction maps into binary strings, (2) detection of overlaps between restriction maps, (3) determining the layout of restriction maps set, (4) creation of consensus genomic maps.
  Our algorithm deals with optical mapping data with low error levels but fails with high-level error reads.
  We developed a software library, console application and module for Python language.
  The approach presented in this paper proved to be faster than a dynamic programming approach and performed well on error-free data.
  It could be used as a step of \textit{de~novo} assembly pipelines or to detect misassemblies.
  The software is freely available in a public repository under GNU LGPL v3 license  (\url{https://sourceforge.net/p/binary-genome-maps/code}).

\end{abstract}

\section{Introduction}

Reducing the cost of sequencing genomes has greatly increased the number of genomic projects and raw read data available.
Better sequencing results have been enabled by better software for assembly and scaffolding.
More efficient algorithms can properly use the increasing volume of available data. With data from different sequencing technologies, Heterogeneous sequencing is an effective method of improving sequencing results.
We propose that optical mapping results be included in heterogeneous sequencing.

Optical Mapping (OM) has been found to be very useful in structural variation analysis, validating genomic assemblies \cite{validation_using_om}, correction and scaffolding assembled contigs \cite{om_scaffolding, howe2015using}.
In the OM process, DNA is first isolated before nicking enzymes are used to digest DNA strands \cite{labeling_om}.
The digested DNA is later imaged \cite{schwartz_om} to determine the distances between the labels on a DNA strand.
The process generates information about the placement of label sites on DNA strands, referred to as restriction maps (rmaps).
Rmaps are analogous to sequence reads in the context of genome sequencing and have a length of 150Kbp -- 1.6Mbp \cite{om_sizes}.
Rmaps can be assembled to create consensus maps.
The result of such an assembly can cover an entire chromosome.
Some of the best-known companies currently developing OM methods are Bionano Genomics \cite{bionano}
and Nabsys \cite{nabsys}.

Other than proprietary software from sequencer manufacturers, few algorithms for creating whole-genome maps have been developed.
Past projects have been based upon dynamic programming with backtracking \cite{computional_om} which scales very poorly with larger genomes, such as the human genome.
Newer algorithms have been developed only for resequencing assembly projects, as they require reference map\cite{omblast, twin}.

In our new algorithm, we explore the possibility of using different genome map representations by binary strings.
Although this algorithm is more sensitive to incorrect indexes of marker positions, it could be used in \textit{de~novo} assemblies, as it does not require a reference genome or reference maps.
Our assembly algorithms could be executed on high-parallel platforms – we have focused on efficiency and the ability to scale with parallel execution.
The software library and application were developed in C++ and are freely available as open-source.

\section{Materials and methods}


Our assembly algorithm is based on an overlap-layout-consensus model.
It uses a binary representation of genomic maps, enabling efficient computational resources.
Short binary sequences can be easily compared using an XOR operation directly implemented in every modern processor.
The algorithm consists of four steps:
\begin{itemize}
  \item Converting rmaps to binary string.
  \item Overlap detection -- scoring the similarities of each pair of rmaps.
  \item Layout forming -- resolving conflicts in neighbouring rmaps.
  \item Consensus -- constructing the contiguous output.
\end{itemize}
The algorithm steps are described below.

\subsection*{Genomic map representation}

\begin{figure}[!ht]
	\includegraphics[width=\columnwidth]{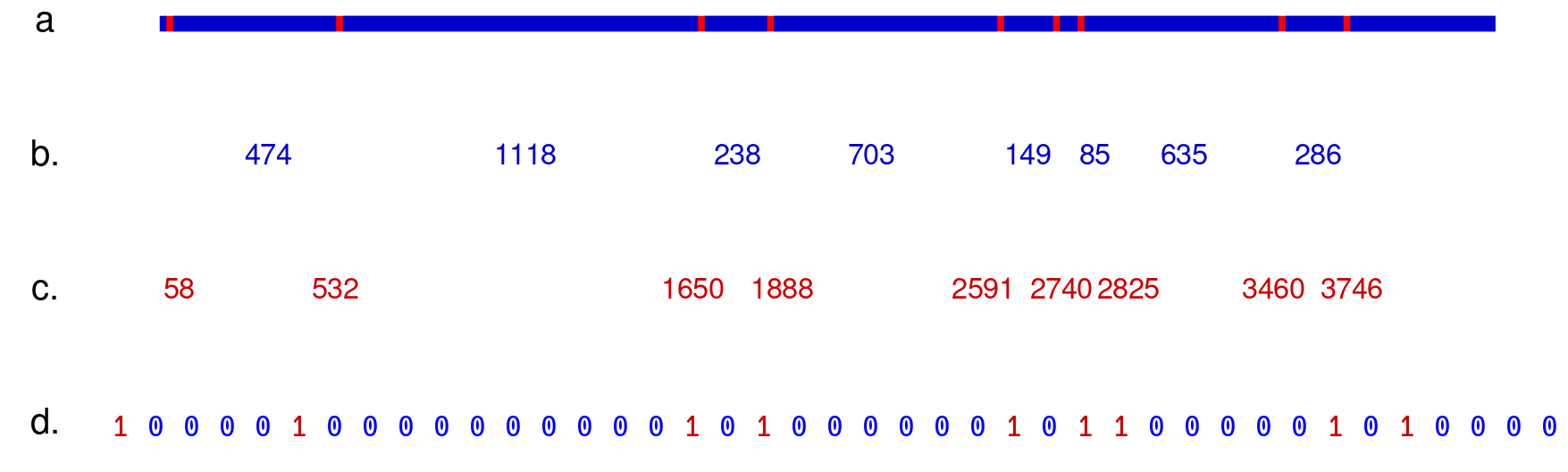}
	\caption{Genome map representations (example for a genome of size 4000). a.-- genome with markers marked in red, b. -- distances between markers, c. -- marker positions, d. -- binary map for quant size = 100.}
	\label{fig:binmap_representation}
\end{figure}

A restriction map (rmap) is a set of marker positions relative to a genome fragment.
A consensus genome map is created during assembly from rmaps for the same genome or chromosome.
The representation of a consensus genome map and a single restriction map is similar. It is an ordered set of marker positions relative to the beginning of a genome fragment or chromosome.
We propose a new representation based on quantization and binary sequences.
Each position in a binary sequence represents a constant genome fragment called a quant, the length of each quant is the same. '1' in the sequence indicates at least one marker present in the quant, while '0' indicates no markers.
We depict such a representation in Figure~\ref{fig:binmap_representation}.

\begin{figure}[!ht]
	\includegraphics[width=\columnwidth]{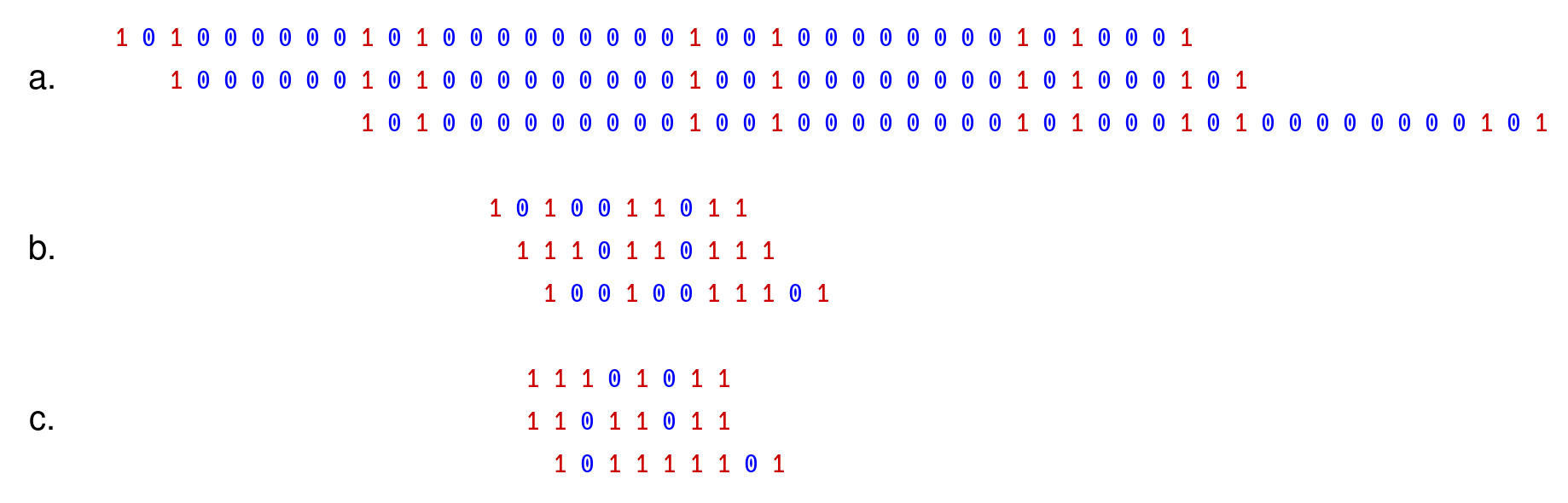}
  	\caption{Quantization errors visualization.
   	a. -- aligned maps from the same genome fragment,
   	b. -- aligned maps from a. quantized with quant length 4,
   	c. -- aligned maps from a. quantized with quant length 5.}
	\label{fig:quantized}
\end{figure}

Converting a rmap into a binary string leads to a loss of information, namely, the exact relative positions of the markers.
We call this issue a~quantization error.
Rmaps from the same genome fragment quantized with the same quant length can produce different binary string representations.
The different placement of quant borders results in the relative placement of markers in binary sequence representations. Moreover, two markers close to each other can be represented as one bit or two neighbour bits.
We present a simple example of such differences in Figure~\ref{fig:quantized}.
These effects are less pronounced with low marker density but must be accounted for in an algorithm.

We assume some level of difference between two rmaps, even if they are \textit{in silico} generated error-free rmaps. However, quantization has been used successfully in genome mapping projects, and quantized maps are still found useful \cite{omblast, kohdista}.

\subsection*{Overlap detection}

After converting rmaps into binary strings, the similarity scores for all pairs of rmaps are calculated.
The similarity scoring is the most time-consuming step due to the high number of rmaps produced in a single experiment.

\begin{figure}[!ht]
	\includegraphics[width=\columnwidth]{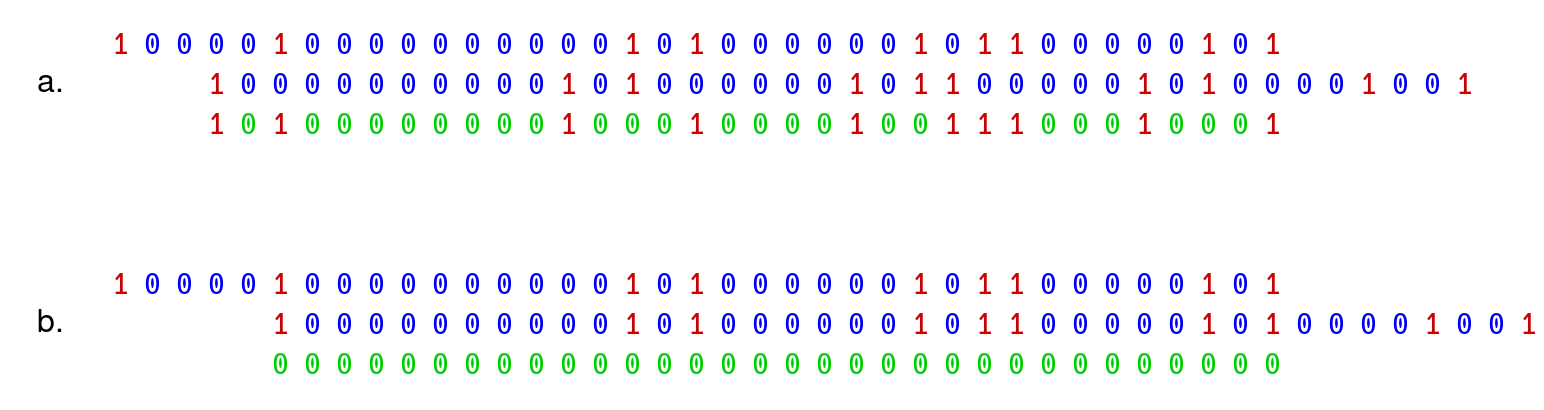}
	\caption{Visualization of two rmaps overlaps and XOR results. In map red indicates marker, blue -- no marker. In results, red indicates difference, green -- no difference. a. -- incorrect alignment, 10 differences; b. -- perfect alignment, no differences.}
	\label{fig:align_xor}
\end{figure}

The binary string representation of rmaps allows a very fast scoring of the similarity of two rmaps in the overlaps detection step.
The algorithm should find the score for the best matching prefix of one rmap to the suffix of the second rmap.
In our approach, for each offsets inside$<-n/2, +n/2>$, n is the length of the binary string the binary exclusive or (XOR) operation is performed.
XOR returns '1' when there is a difference at a given position, while '0' indicate confirmation, as depicted in Figure~\ref{fig:align_xor}.
The XOR operation is implemented as a single instruction between registers in all general-purpose CPUs.
Therefore, our algorithm is very fast because it uses operations implemented directly in the hardware.
The number of differences (number of '1' returned by XOR) determines a map’s similarity for a given offset.
We present our approach to overlap detection in Algorithm~\ref{alg:overlap}.

\begin{algorithm}[!ht]
	\caption{Overlap search for left neighbor}
	\begin{algorithmic}[1]
		\Function{findLeftAlignment}{ref, aligned}
		\State $ maxShift \gets \Call{minLen}{ref, aligned}$
		\Statex \Comment{could be adjusted to be e.g. half of smaller rmap}
		\State $ bestAlign \gets 1$
		\Statex \Comment{differenting part, 1 means that all bits differ, 0 means full conformance.}
		\State $ bestShift \gets 0 $
		\Statex \Comment{shift between rmaps ends}
		\For{$shift \in \{1, ..., maxShift\}$}
		\State $test \gets aligned \ll shift$
		\State $result \gets test \textsc{ xor } ref$
		\State $\Call{truncateLonger}{test, ref}$
		\If{$\Call{count}{result}$ / $\Call{len}{result} < bestAlign$} 			\Statex\Comment{it's a better alignment}
		\State $ bestAlign \gets \Call{count}{result}$ / $\Call{len}{result}$
		\State $ bestShift \gets shift $
		\EndIf
		\EndFor

		\Return{$bestShift$}
		\EndFunction
	\end{algorithmic}
	\label{alg:overlap}
\end{algorithm}

\subsection*{Layout forming}

Our analysis of multiple genomes and enzymes shows that the markers density is similar. As mentioned in the 'Genomic map representation' subsection, a binary representation loses information. In an extreme case (quant length is too big), the binary string is '11...1' (only '1').
We assume that the binary string represents a given rmap, so it has at most 10\% of '1'.
This assumption could be easily fulfilled - we could decrease the quant size if we had too much 1.
Using this assumption, we can consider in layout formatting only rmaps with differences below a given threshold and with overlaps above a given length.
For example, if maps have an average density of markers around 10\%, rmaps aligned with more than 10\% of differences can be discarded.
We selected the threshold of 10\% experimentally so that the length of binary strings did not increase the computation time too much. Our algorithm has a parameter that allows the end-user to set another value. We commonly set this value to 20\%, or even 30\% in numerical experiments, also allowing less precise alignments.

Our method for layout determination is presented as Algorithm~\ref{alg:layout}.

\begin{algorithm}[!ht]
	\caption{Global rmaps alignment}
	\begin{algorithmic}[1]
		\Procedure{findGlobalAlignment}{rmaps}
		\For{$i \in \{0,1... \Call{len}{rmaps}\}$}
		\State $smallestShift \gets \Call{len}{rmaps[i]}$
		\State $leftRMap \gets i$
		\For{$j \in \{0,1... \Call{len}{rmaps}\}$}
		\If{$i == j$}
		\State \textbf{continue}
		\EndIf
		\State $result \gets \Call{findLeftAlignment}{rmaps[i], rmaps[j]} $
		\If{$smallestShift >result $}
		\State $smallestShift \gets result$
		\State $leftRMap \gets j$
		\EndIf

		\EndFor
		\State $rmaps[i].addLeftNeighbor(leftRMap)$
		\EndFor
		\EndProcedure
	\end{algorithmic}
	\label{alg:layout}
\end{algorithm}

\begin{figure}[!ht]
	\includegraphics[width=\columnwidth]{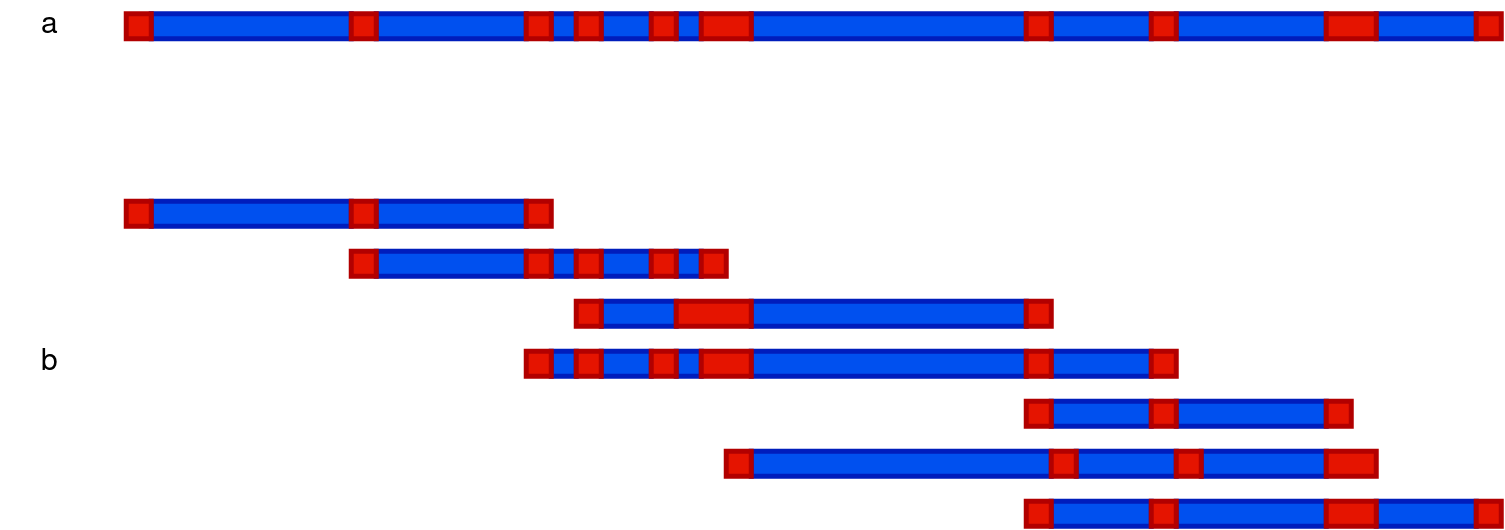}
	\caption{Visualization of aligned rmaps.
	Red indicates marker in map, blue indicates no marker.
	a. -- reference map,
  	b. -- set of aligned rmaps.}
	\label{fig:aligned_rmaps}
\end{figure}

After rmaps are detected as neighbours we prepare them for alignment.
We eliminate rmaps containing themselves entirely in others.
We present a visualization of rmaps layouts in Figure~\ref{fig:aligned_rmaps}.

\subsection*{Consensus map creation}

Contiguous sequences of neighbour rmaps are constructed after aligning all the maps. 
For contigs detection, we interrogate the graph, where rmaps are the vertex, and the similarity score is the edge.
If several vertices form a line, these vertices are traversed using a consensus algorithm.
If several vertices form a circle on a graph, the circle is broken.

Each sequence of rmaps is traversed using a consensus algorithm to create a contiguous map representation.
For each position on the aligned rmaps, the number of '1' and '0' are counted, then the most frequent symbol is chosen.
In the case of an equal number of '1' and '0', the symbol '1' is chosen because the rmap representation is sparse and losing information about a marker is considered a more significant issue than adding a non-existing marker.
We present this process in Algorithm~\ref{alg:consensus}.

\begin{algorithm}[!ht]
	\caption{Consensus algorithm}
	\begin{algorithmic}[1]
		\Function{performConsensus}{contig}
		\State $ map \gets $ [ ]
		\State $ count \gets 0 $
		\For{$p \in \{0,1, ..., \Call{len}{contig} - 1 \} $}
		\Statex \Comment{\textit{p} - position from beginning of overlapping maps}
		\For{$rmap \in \Call{mapsAtPos}{contig, p}$}
		\If{$rmap.at(p) == 1 $} \Comment{rmap.at(p) - value on position \textit{p}}
		\State $ count \gets count + 1 $
		\EndIf
		\EndFor
		\If{$count < (k / 2) $}
		\State $ map.append(0) $
		\Else
		\State $ map.append(1) $
		\EndIf
		\EndFor

		\Return{$map$}
		\EndFunction
	\end{algorithmic}
	\label{alg:consensus}
\end{algorithm}

\subsection*{Assembly parameters}

Our application allows users to adjust certain parameters of the assembly:
\begin{itemize}
	\item \textit{margin quant} -- the minimal length of overlap (i. e., common fragment of rmaps pair) in quants.
	\item \textit{margin part} -- minimal part of the smaller map to consider for overlap (i. e. if the margin is 0.25, the overlap has to be at least 25\% of smaller map length in size)
	\item \textit{threshold} -- the threshold of differences to consider an overlap meaningful (i. e. if the threshold is 0.15, overlaps with more than 15\% of differences on the common part are discarded)
	\item \textit{marker preference} -- consensus preference, symbol to be used in resulting contig if there are 50\% of both '0' and '1' 
	\item \textit{folds} -- how many times assembly step has to be performed before giving output.
\end{itemize}

Users can also adjust input and output filenames and the number of threads used. 
BGM is open source, and the assembly process is modular.
For example, further adjustments can be made by changing the cost function in the source code.

\section{Results}

We performed numerical experiments on simulated and real datasets.
We compared our approach with an algorithm based on dynamic programming with backtracking.
Moreover, we analyzed the computational complexity of our new algorithm. In our experiments we used data from three organisms : \textit{Escherichia coli} \cite{ecolidata}, \textit{Caenorhabditis elegans} \cite{roundwormdata}, and \textit{Acinetobacter baumannii} \cite{realdata}. We provide a brief genome sizes summary of the species used in Table~\ref{tab:species}.

\begin{table}[!ht]
    \caption{Summary of genome sizes and chromosomes count.}
	\centering
    \begin{tabular}{ccc}
        \hline
        species      & chromosomes & length {[}Mbp{]} \\ \hline
        \textit{E.~coli}      & 1           & 4.6              \\
        \textit{C.~elegans}   & 7           & 100.2            \\
        \textit{A.~baumannii} & 1           & 3.8             \\ \hline
    \end{tabular}
    \label{tab:species}
\end{table}

\subsection*{Simulated data}

\begin{figure}[!ht]
	\includegraphics[width=\columnwidth]{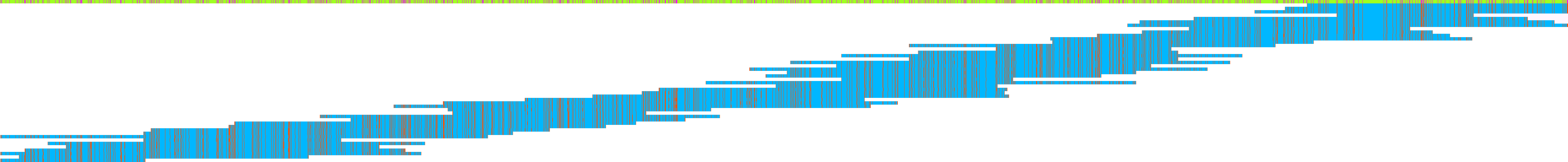}
  	\caption{Visualization of simulated rmaps dataset aligned to a reference genome. The reference genome is marked in green, the rmaps in blue and markers are denoted with red. Rmaps come from the \textit{E.~coli} genome for the \bspq{} enzyme and $15\times$ coverage.}
	\label{fig:e.coli_rmaps_x15}
\end{figure}

We simulated error-free rmaps using a reference genome.
The rmaps length was sampled with the normal distribution of a given mean and standard deviation, and the rmaps position was sampled using a uniform distribution.
Marker positions were then calculated, with fragments on the ends bearing no markers being removed.
Finally, the rmap was quantized with a given quant value (1000 bp in our case).
An example of a simulated dataset aligned to a reference genome is presented in Figure~\ref{fig:e.coli_rmaps_x15}.

Using simulated datasets from the \textit{E.~coli} genome \cite{ecolidata} we reconstructed the whole genome consensus map, with simulated \bspq{} marker and $7.5\times$ coverage of the genome.
For \bsss{} markers we reconstructed the whole genome for $10\times$ coverage.
Both datasets are summarized in~Table~\ref{tab:e.coli_experiment}.

\begin{table}[!ht]
	\caption{Input data that enabled reconstruction of a whole \textit{E.~coli} reference map.}
	\centering
	\begin{tabular}{cccc}
		\hline
		enzyme & coverage & rmap count & avg length [Kbp] \\ \hline
		\bspq{} & 7.5      & 36          & 959.30                \\
		\bsss{} & 10       & 47          & 1093.00               \\ \hline
	\end{tabular}
	\label{tab:e.coli_experiment}
\end{table}

\begin{figure}[!ht]
	\includegraphics[width=\columnwidth]{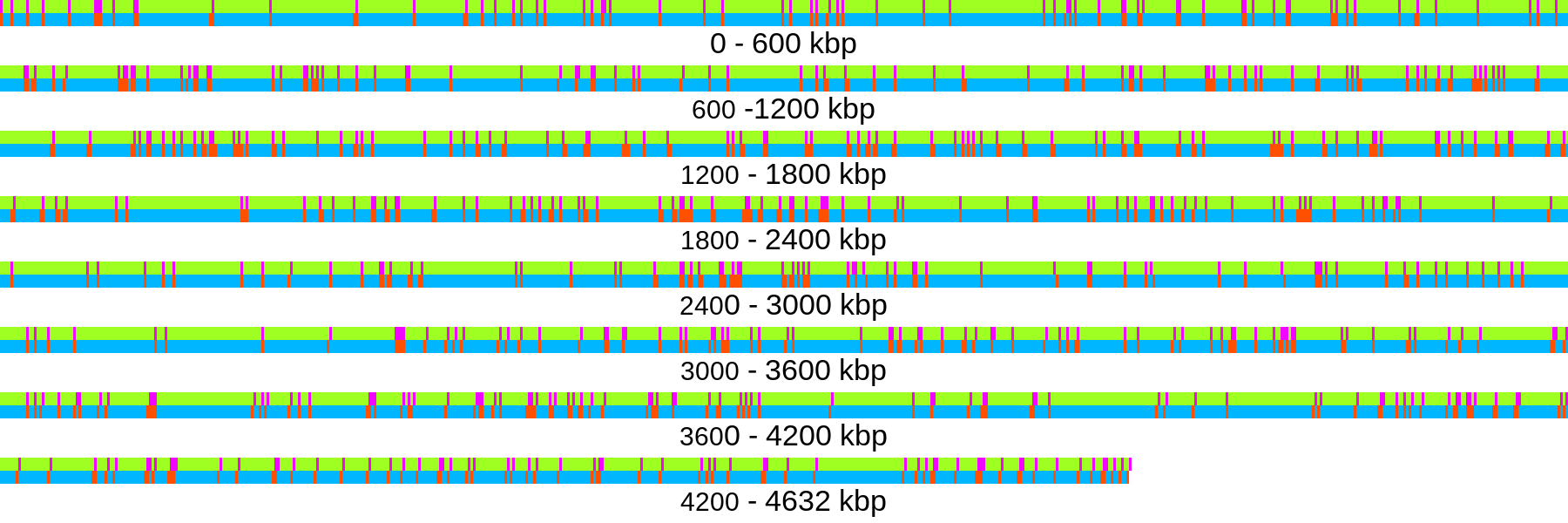}
  	\caption{Visualization of assembly results for simulated \textit{E.~coli} dataset for \bspq{} simulated enzyme with $7.5\times$ coverage. Reference genome is marked in green, result in blue and markers are denoted by red.}
	\label{fig:e.coli_result_split}
\end{figure}

The the experiment result is visualized in~Figure~\ref{fig:e.coli_result_split}.

\subsection*{Tuning cost function for overlapping}

We analyzed the best overlap using two different cost functions:
\begin{enumerate}
  \item Number of differences on the overlapping fragment.
  \item Differentiating level calculated as the number of differences divided by overlap length.
\end{enumerate}

For comparison, we generated four datasets.
Rmap length was sampled from a normal distribution with an expected value of 1 Mbp and a standard deviation of 300 Kbp.
The datasets statistics are summarized in Table~\ref{tab:input_cost_f}.
We only changed the cost function during our experiments, leaving the remaining parameters unaltered.
The parameters used in this experiment were : \textit{margin\_part}~=~0.5, \textit{margin\_quant}~=~100, \textit{folds}~=~2, \textit{thresh}~=~0.1.
The comparison was run on simulated datasets from \textit{E.~coli} \cite{ecolidata} and \textit{C.~elegans} \cite{roundwormdata} reference genomes.
The results of our experiments are also summarized and presented in Table~\ref{tab:result_cost_f}.

\begin{table}[!ht]
	\caption{Input data for cost function experiments}
	\begin{tabular}{ccccc}
		\hline
		genome  & enzyme & coverage & rmap count & avg. rmap length [Kbp] \\ \hline
		\textit{E.~coli} & \bspq{} & 10       & 47          & 983.38                \\
        \textit{E.~coli} & \bspq{} & 20       & 92          & 1003.08               \\
		\textit{C.~elegans} & \bsss{} & 10       & 967         & 1031.72               \\
		\textit{C.~elegans} & \bsss{} & 40       & 3943        & 1008.60               \\

    \hline
	\end{tabular}
	\label{tab:input_cost_f}
\end{table}

\begin{table}[!ht]
	\caption{Results of cost function experiments}
	\begin{tabular}{cccccc}
		\hline
		function & dataset & contigs & avg. length  & N50   & longest contig  \\
		  &   &  count  & [Kbp]        & [Kbp] & [Kbp]           \\ \hline
		1       & \textit{E.~coli}, \bspq{} 10x & 2                         & 3167                      & 4212  & 4212                      \\
		2       & \textit{E.~coli}, \bspq{} 10x & 2                         & 3387                      & 4633  & 4633                      \\
		1       & \textit{E.~coli}, \bspq{} 20x & 3                         & 2748                      & 4633  & 4633                      \\
		2       & \textit{E.~coli}, \bspq{} 20x & 3                         & 2497                      & 4633  & 4633                      \\
		1       & \textit{C.~elegans}, \bsss{} 10x & 79                        & 2982                      & 2413 & 8436                      \\
		2       & \textit{C.~elegans}, \bsss{} 10x & 80                        & 3056                      & 2375 & 7242                      \\
		1       & \textit{C.~elegans}, \bsss{} 40x & 38                        & 11984                     & 5263 & 17718                     \\
		2       & \textit{C.~elegans}, \bsss{} 40x & 41                        & 11221                     & 4985 & 20769                     \\ \hline
	\end{tabular}
	\label{tab:result_cost_f}
\end{table}

All rmaps were aligned with at least one other.
After analyzing the results, we found that the second function, when the cost was calculated as the number of differences divided by the overlap length, allowed us to connect rmaps more effectively.
We decided to use this function in further experiments.

\subsection*{Comparing results of our algorithm with dynamic programming approach}

We compared our algorithm to a dynamic programming approach with backtracking \cite{valouev_align, valouev_asm}.
We ran both algorithms on a single thread for comparison purposes, and parameters were unchanged for all runs.
We used a server with a Xeon E5-2637 v2 processor (2 cores 4 threads, 3.5 GHz \cite{arkIntel}).
The comparison was run on simulated datasets from \textit{E.~coli} \cite{ecolidata} and \textit{C.~elegans} \cite{roundwormdata} reference genomes.
We compared performance and quality, measured as run times and the number of contigs.
We ran software with a dynamic programming approach developed by \textit{Valouev et al} \cite{valouev_ref}.

\begin{figure}[!ht]
  \includegraphics[width=\columnwidth]{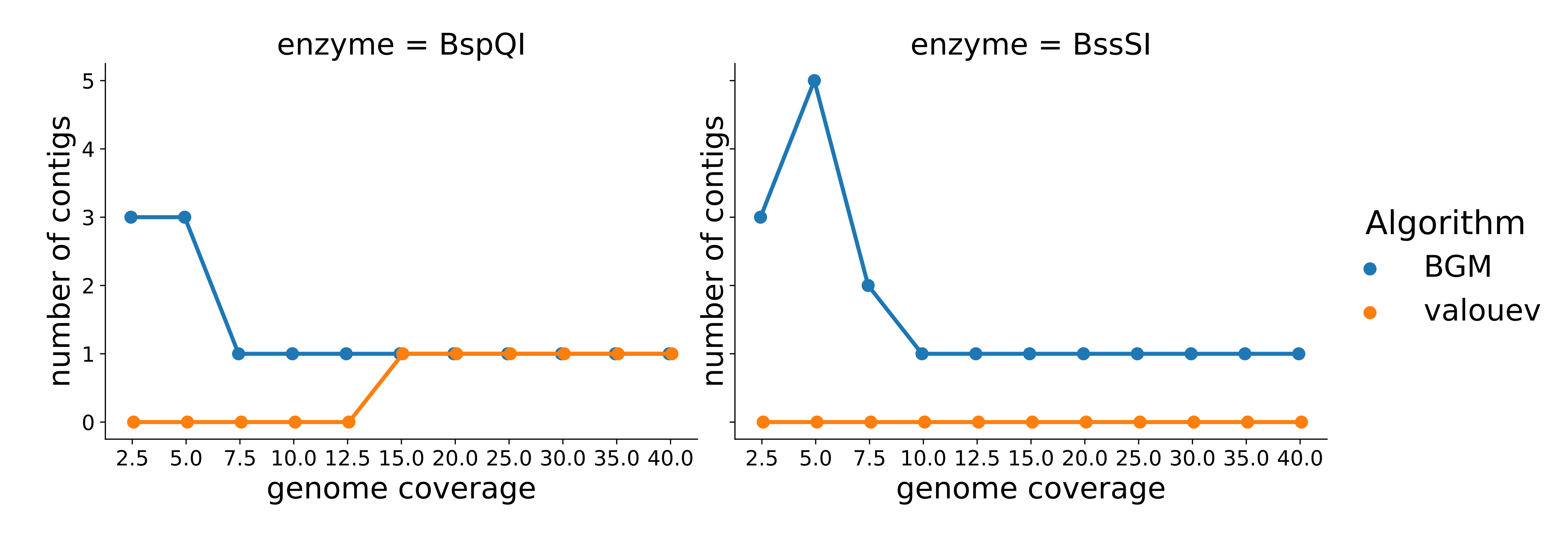}

  \includegraphics[width=0.55\columnwidth]{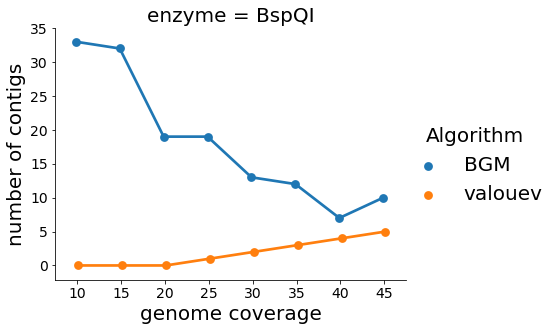}
  \caption{Comparison of contigs count depending on coverage of the genome, \textit{E.~coli} dataset (on top), \textit{C.~elegans} dataset (on bottom).}
  \label{fig:compare_contigs}
\end{figure}

Figure~\ref{fig:compare_contigs} summarizes the contigs produced by both algorithms on the \textit{E.~coli} dataset.
We assume rmaps without gaps cover the input genome (Lander-Waterman theory \cite{lander1988genomic} for genomes used, coverages and rmap sizes).
Therefore, we measure the quality of alignment as a number of contigs. The smaller the number of contigs, the better.

The dynamic programming approach needed twice as much coverage to produce a contig with \bspq{} than our approach did.
When simulating \bsss{} markers, our algorithm reconstructed the whole genome map with $10\times$ coverage,
whereas the dynamic programming approach did not produce any contig at all, even with $40\times$ coverage.
Due to this, we simulated only the \bspq{} enzyme when comparing algorithms on the \textit{C.~elegans} genome \cite{roundwormdata}.
A summary of the contigs counts during our experiments with the simulated \textit{C.~elegans} dataset is presented in Figure~\ref{fig:compare_contigs}.

Our experiments show that an algorithm based on binary strings can reconstruct a whole genome with less coverage.
The parameters need to be tuned to a specific dataset for best results.

\begin{figure}[!ht]
  \includegraphics[width=0.49 \columnwidth]{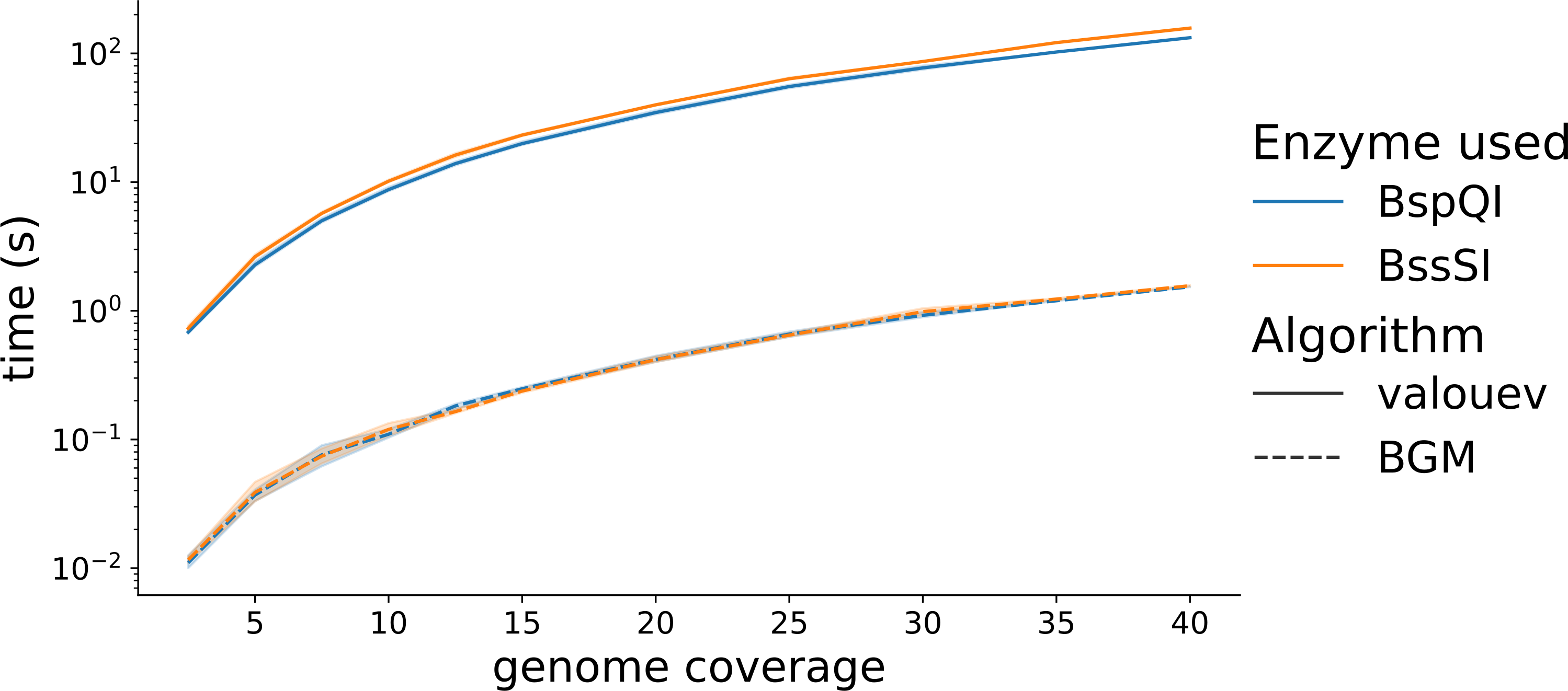}
  \includegraphics[width=0.49 \columnwidth]{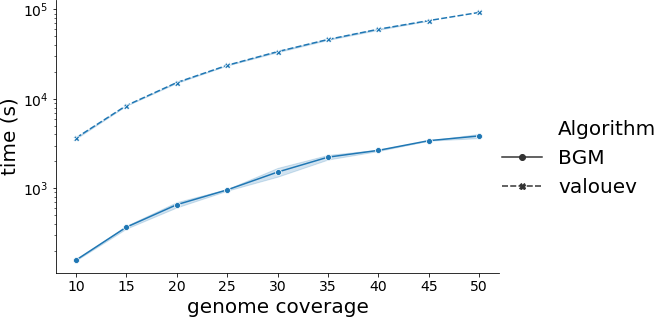}

	\caption{Comparison of time needed for algorithms to finish assembly depending on genome coverage of \textit{E.~coli} dataset (on left), and \textit{C.~elegants} dataset (on right).}
	\label{fig:compare_ecoli_times}
\end{figure}

We measured the time needed for both algorithms to finish the assembly.
The results are summarized in Figure~\ref{fig:compare_ecoli_times}.
From all runs, we can see that our approach delivered results around 80 times faster than the dynamic approach on the datasets tested.
The complexity of both algorithms is \( \Omega(n^2)\) because each rmap must be analyzed individually, and in both cases, the algorithm must seek an overlap with another rmap.

\subsection{Experiments with exponential rmaps length distribution}

We performed an additional experiment using datasets where length was generated using exponential distribution with parameter \(\beta = 150 kbp \)\footnote{\(\beta\) is the scale parameter, which is the inverse of the rate parameter \( \beta = 1 / \lambda \). The rate parameter is an alternative, widely used parameterization of the exponential distribution.} Maps shorter than 100 kbp were discarded, resulting in 275 kbp average length of an rmap. Datasets used were summarized in Table~\ref{tab:exponential_summary}.

\begin{table}[!ht]
	\caption{Summary of exponential length datasets}
	\begin{tabular}{cccc}
		\hline
        enzyme & coverage & map count & avg. length {[}kbp{]} \\
        \hline
        \bsss{}  & 10       & 173       & 258.73                \\
        \bsss{}  & 20       & 320       & 279.92                \\
        \bsss{}  & 30       & 484       & 277.45                \\
        \bsss{}  & 40       & 651       & 274.58                \\
        \bsss{}  & 60       & 965       & 277.94                \\
        \bspq{}  & 10       & 170       & 261.48                \\
        \bspq{}  & 20       & 316       & 281.67                \\
        \bspq{}  & 30       & 480       & 277.30                \\
        \bspq{}  & 40       & 643       & 276.26                \\
        \bspq{}  & 60       & 957       & 278.38               \\
        \hline
    \end{tabular}
    \label{tab:exponential_summary}
\end{table}

We were able to reconstruct the whole \textit{E.~coli} genome map using $30\times$ coverage, both with \bspq{} and \bsss{} enzymes simulation.
We obtained whole genome maps using parameters :  \textit{margin\_part}~=~0.3, \textit{margin\_quant}~=~90, \textit{thresh}~=~0.1. \bspq{} dataset required \textit{folds} parameter to be set on 3 or more, where with \bsss{} 2 was enough.
The dynamic programming approach was able to reconstruct the genome only with $60\times$ coverage and \bspq{} enzyme. There was also a partial contig reconstructed with $60\times$ coverage and \bsss{} enzyme. Apart from that, the dynamic programming approach produced no contigs.

\subsection*{Experiments with real data}

We performed experiments using a real dataset from the optical mapping of the \textit{A.~baumannii} genome \cite{realdata}.
Rmaps from this dataset were converted into binary form using segment lengths of 1000 bp and 1500 bp.
We used the entire dataset and subset containing only maps longer than 300 Kbp in both cases.
A summary of the datasets is presented in Table~\ref{tab:input_real}.

We could not create a consensus map with these datasets, despite changing the parameters.
For analysis, we used results from the original genomic project \cite{realdata} because the authors did not provide the information about the enzyme used.
The statistics of the two experiments on filtered datasets are presented in Table~\ref{tab:results_real}.

\begin{table}[!h]
	\caption{Summary of real datasets from \textit{A.~baumannii} used in experiments.}
	\begin{tabular}{ccc}
		\hline
		dataset                    & rmap count & avg rmap length [Kbp] \\ \hline
		all                 & 51389       & 214.67                \\
		rmapy longer than 300 Kbp & 7359        & 385.20                \\ \hline
	\end{tabular}
	\label{tab:input_real}
\end{table}

\begin{table}[!h]
    \caption{Summary of experiments with real dataset from \textit{A.~baumannii}.}
	\begin{tabular}{ccccccc}
		\hline
		margin\_part & margin\_quant & thresh & folds & \begin{tabular}[c]{@{}l@{}}contigs \\ count\end{tabular} & \multicolumn{1}{l}{\begin{tabular}[c]{@{}l@{}}N50\\ {[}Kbp{]}\end{tabular}} & \begin{tabular}[c]{@{}l@{}}avg contig \\ length {[}Kbp{]}\end{tabular} \\ \hline
		0.5          & 65            & 0.08   & 2     & 784                        & 395           & 409.41                     \\
		0.7          & 65            & 0.065  & 1     & 1176                       & 408           & 426.99                     \\ \hline
	\end{tabular}
	\label{tab:results_real}
\end{table}

\begin{figure}[!ht]
	\includegraphics[width=\columnwidth]{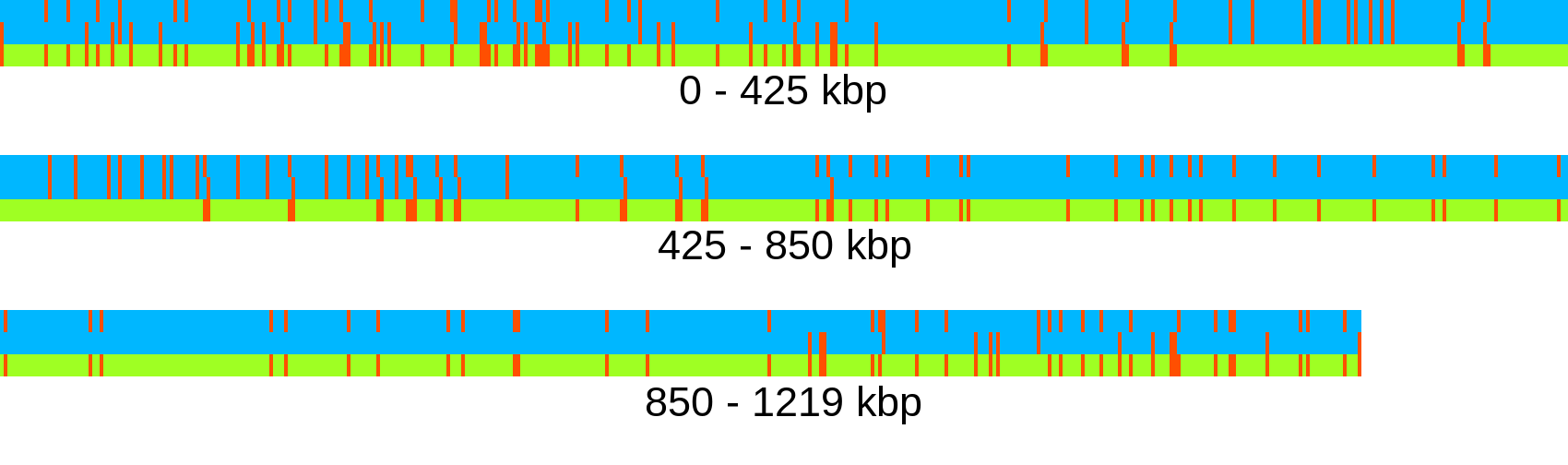}
	\caption{Example of contig with empty space in centre. The reference map and contig are presented in blue with red markers. The reference map is at the top, the contig in the middle. The last row is a comparison where red signifies difference and green signifies conformity.}
	\label{fig:empty_center}
\end{figure}

\begin{figure}[!ht]
	\includegraphics[width=\columnwidth]{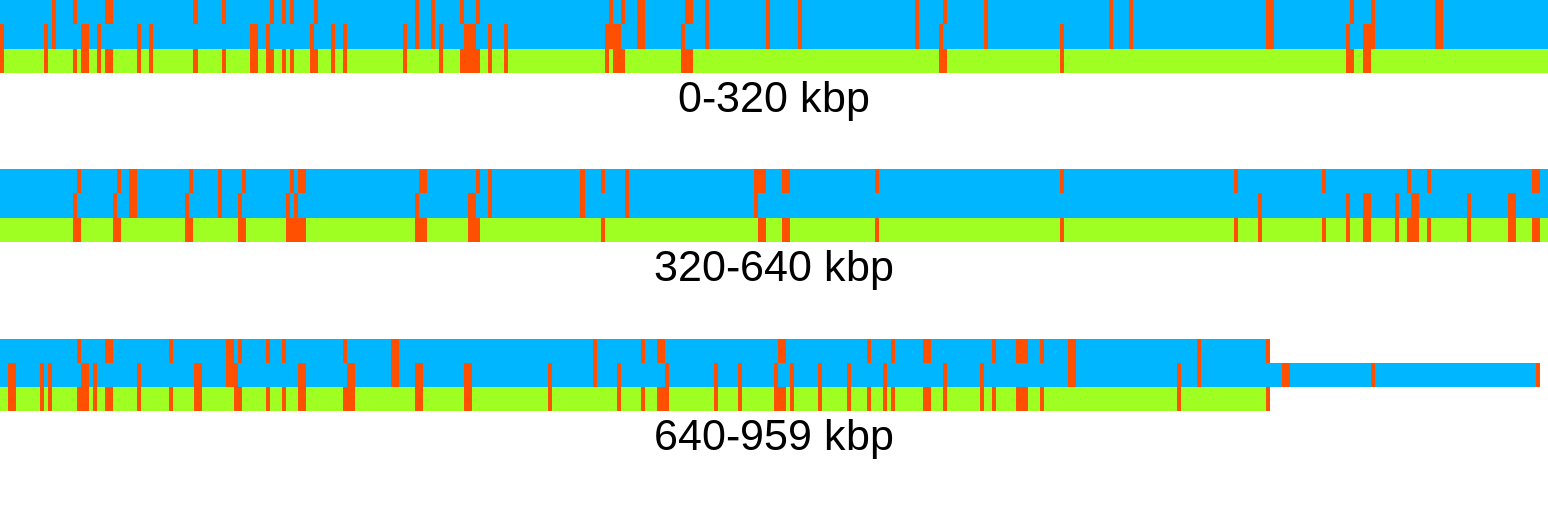}
	\caption{Example of contig with errors at its ends. The reference map and contig are presented in blue with red markers. The reference map is at the top and the contig in the middle. The last row is a comparison where red signifies difference and green signifies conformity.}
	\label{fig:error_ends}
\end{figure}

Even though the dataset has around $725\times$ genome coverage, we could not create valid contigs.
During the experiments, we encountered two anomalies that frequently occurred on the resulting contigs.
Some of the maps had long areas free from any markers not present on the reference map.
An example of such a contig is presented in Figure~\ref{fig:empty_center}.
The other anomaly in the results was a significant increase in error frequency the longer the distance from the correct fragment.
An example of such an effect is presented in Figure~\ref{fig:error_ends}.

We assume that most errors were caused by sizing errors and different stretch factors of the molecules scanned.
Our algorithm did not address such errors due to the nature of binary sequences.

\subsection*{Experiment on simulated \textit{A.~baumannii}}

We performed experiments using a simulated dataset from the same species to check if the genome sequences were not the source of the problems encountered during our experiments with real datasets.
We used the reference genome of the bacteria\cite{bannumaniidata} and generated rmaps datasets simulating nicking of \bspq{} and \bsss{} enzymes.
The length was sampled from a normal distribution with an expected value of 400 Kbp and a standard deviation of 150 Kbp to match map lengths from a filtered real dataset.
We summarize the datasets created in Table \ref{tab:input_verify}.
For parameters \textit{margin\_part}~=~0.3, \textit{margin\_quant}~=~105, \textit{folds}~=~2, \textit{thresh}~=~0.19. For rmaps generated simulating the \bspq{} enzyme we were able to obtain a consensus map covering the whole genome.
For rmaps generated using the enzyme, greater sequencing coverage was necessary, and the \textit{folds} parameter \bsss{} was increased to $3$.
We were also able to create whole-genome consensus maps using datasets with higher (i. e. $40-70\times$) genome coverage.

\begin{table}[!ht]
	\caption{Summary of simulated datasets from \textit{A.~baumannii} used in experiments.}
	\begin{tabular}{cccc}
		\hline
		enzyme & coverage & rmaps count & average length [Kbp] \\ \hline
		\bspq{} & 25       & 244         & 391.53                \\
		\bsss{} & 25       & 244         & 397.07                \\ \hline
	\end{tabular}
	\label{tab:input_verify}
\end{table}

\section{Discussion}

In our experiments, we were able to reconstruct the genomes of \textit{E.~coli} and \textit{C.~elegans} from simulated error-free datasets.
Our algorithm proved faster than a dynamic programming approach and produced results with less genome coverage needed.
We used two different enzymes \bspq{} and \bsss{} to show that algorithm results are similar for both cases. Our algorithm should work correctly for any enzyme if the sites are spread over the genome.

Our approach can deal with quantization artefacts and the inaccuracy of single-site placement. BGM can also deal with OM errors like false and missing sites, which are present in a binary string as a flipped bit '1' instead of  '0' and '0' instead of '1', respectively.
Our algorithms deal with dose mistakes and artefacts by allowing differences between rmap alignments. When noise is more significant than a threshold (this is an algorithm parameter set by the user), two noisy rmaps will not be connected.
However, due to the nature of binary strings, this approach cannot deal with other kinds of errors, resulting in insertion, deletion, or shift in a binary string. Examples of such errors are stretch variance, sizing errors and missing fragments.

BGM creates a consensus restriction map on chromosome length and could be used to find structural rearrangements.
Moreover, BGM can be used as a hybrid de novo assembly step.
As presented in our previous work \cite{rn:sdata2019hymenolepis} a~simultaneous use of available sequencing technologies is useful in improving genome assemblies.
BGM algorithm deals with optical mapping data with low error level, e.g. provided by nabsys \cite{oliver2017high}.

With real datasets, our algorithm struggled to produce valid sequences.
In our opinion, this effect is due to sizing errors present in the optical mapping data.
Even a slightly stretched map of the same fragment of the genome could not be aligned perfectly.

Our approach could be modified and used to align contigs on consensus maps as both contigs and consensus maps have a low error rate.
Another option is to use correction algorithms \cite{cor_err_comet, cor_err_elmeri} focused on resolving to size and stretch errors as a part of the assembly pipeline.
BGM is fully open source and available at \url{https://sourceforge.net/projects/binary-genome-maps/}.
We would be grateful if someone could adopt any correction algorithm or create a new one.
We also share all of the datasets mentioned and encourage the community to perform experiments using our software.
We are open to all contributions to our project.

\section{Conclusion}

As more genome mapping and heterogeneous genomic projects are created, it is crucial to develop open-source algorithms to deal with such data.
To this end, our paper presents an algorithm based on binary representation to create consensus maps from rmaps.
However, judging from the results of the experiments, this algorithm is, in our opinion, better suited for applications with error-free maps like contigs alignment.

\section*{Availability and requirements}

BGM is implemented in C++, and is freely available under GNU Library or Lesser General Public License version
3.0 (LGPLv3). It and related materials (like all datasets used) can be downloaded from project homepage \url{https://sourceforge.net/projects/binary-genome-maps/}.

\section*{Competing interests}
The~author declares that he has no competing interests.

\section*{Author contributions}

PS and RN Identified the problem and designed the approach.,
PS Implemented the software, worked on testing and validation.
PS Wrote the main manuscript text and prepared figures.
All authors read and approved the final manuscript.

\section*{Acknowledgements}

This research was funded by Warsaw University of Technology grant CyberiADa-1.

\bibliographystyle{abbrv}
\bibliography{om.bib}

\end{document}